\documentclass[5p, 10pt, sort&compress]{elsarticle}

\usepackage{amsmath,amssymb,graphicx,enumerate,multirow,array,ragged2e,siunitx}
\usepackage[colorlinks=true,breaklinks=true]{hyperref}
\makeatletter
\def\ps@pprintTitle{%
 \let\@oddhead\@empty
 \let\@evenhead\@empty
 \def\@oddfoot{}%
 \let\@evenfoot\@oddfoot}
\makeatother

\usepackage[caption=false]{subfig}

\usepackage[dvipsnames]{xcolor}
\newcommand{\revision}[1]{{\color{black}{#1}}}

\definecolor{revisiontwo}{rgb}{0,0,0}

\usepackage{framed} % Framing content
\usepackage{multicol} % Multiple columns environment
\usepackage{nomencl} % Nomenclature package
\makenomenclature
\renewcommand*\nompreamble{\begin{multicols}{2}}
\renewcommand*\nompostamble{\end{multicols}}
\DeclareMathSizes{10}{9}{7}{5}
\begin{document}
%\small

\begin{frontmatter}

\title{Rear-surface integral method for calculating thermal diffusivity:\\ finite pulse time correction and two-layer samples}

\author[qut]{Elliot J. Carr}
\ead{elliot.carr@qut.edu.au}

\author[qut]{Christyn J. Wood}
%\ead{}

\address[qut]{School of Mathematical Sciences, Queensland University of Technology (QUT), Brisbane, Australia.}

\journal{International Journal of Heat and Mass Transfer}
%\journal{Chemical Engineering Science}


\begin{abstract}
We study methods for calculating the thermal diffusivity of solids from laser flash experiments. This experiment involves subjecting the front surface of a small sample of the material to a heat pulse and recording the resulting temperature rise on the opposite (rear) surface. Recently, a method was developed for calculating the thermal diffusivity from the rear-surface temperature rise, which was shown to produce improved estimates compared with the commonly used half-time approach. This so-called \textit{rear-surface integral method} produced a formula for calculating the thermal diffusivity of homogeneous samples under the assumption that the heat pulse is instantaneously absorbed uniformly into a thin layer at the front surface. In this paper, we show how the rear-surface integral method can be applied to a more physically realistic heat flow model involving the actual heat pulse shape from the laser flash experiment. New thermal diffusivity formulas are derived for handling arbitrary pulse shapes for either a homogeneous sample or a heterogeneous sample comprising two layers of different materials. Presented numerical experiments confirm the accuracy of the new formulas and demonstrate how they can be applied to the kinds of experimental data arising from the laser flash experiment.
\end{abstract}


\end{frontmatter}

\section{Introduction}
\begin{figure*}[t]
\centering
\includegraphics[width=1.0\textwidth]{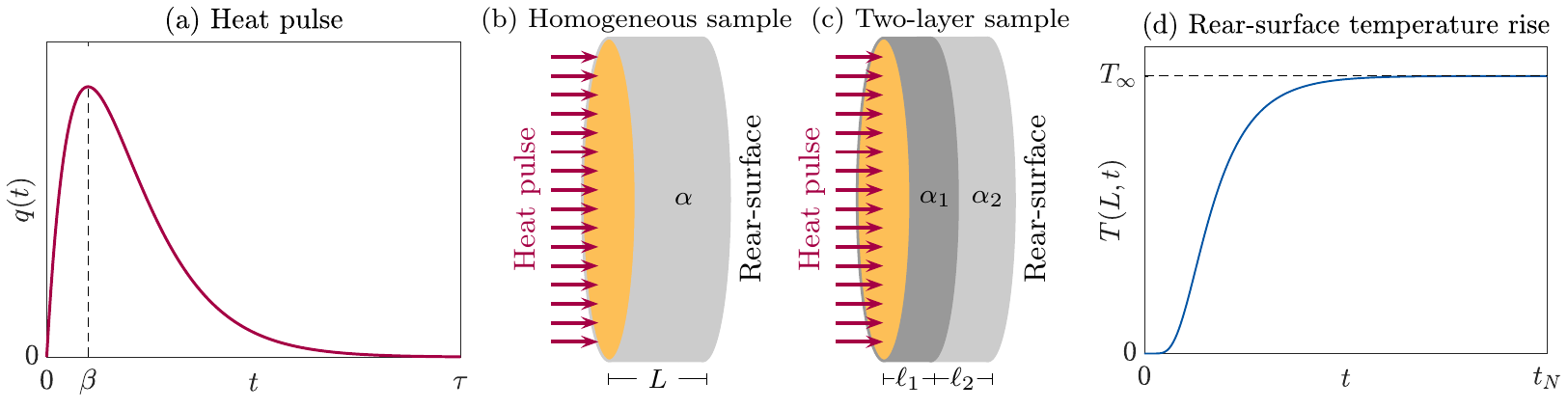} 
\caption{Schematic representation of the laser flash experiment for (b) a homogeneous sample with thickness $L$ and thermal diffusivity $\alpha$ and (c) a two-layer sample consisting of two homogeneous layers of thickness $\ell_{1}$ and $\ell_{2}$ and thermal diffusivities $\alpha_{1}$ and $\alpha_{2}$. A heat pulse of radiant energy is applied to the front surface (a) resulting in the temperature on the opposite (rear surface) of the sample rising over time (d). Eventually the temperature rise tends to a finite steady-state value, $T_{\infty}$, under the assumption of a thermally insulated sample.}
\label{fig:laser_flash}
\end{figure*}

The most popular technique for measuring the thermal diffusivity of a solid material is the laser flash method \citep{blumm_2002,czel_2013,vozar_2003}. Originally developed by \citet{parker_1961}, this method involves subjecting the front surface of a small sample of the material to a heat pulse of radiant energy and recording the resulting temperature rise on the opposite (rear) surface of the sample (Figure \ref{fig:laser_flash}). The thermal diffusivity is then calculated from the rear-surface temperature rise curve by developing a mathematical model describing the heat flow emanating from the front surface and analysing the solution of this model at the rear surface. 

\citet{parker_1961} assumed the sample is homogeneous and thermally insulated; the heat pulse is instantaneously and uniformly absorbed by a thin layer at the front surface; and the heat flow is one-dimensional in the direction of the thickness of the sample extending from the front surface to the rear surface (Figure \ref{fig:laser_flash}b). These simplifying assumptions yield the now famous and widely-used formula \cite{jeon_2019,wang_2017,zajas_2013,zhao_2016} for the thermal diffusivity 
\begin{gather}
\label{eq:alpha_parker}
\alpha \approx \frac{1.37L^{2}}{\pi^{2}t_{0.5}},
\end{gather}
where $L$ is the thickness of the sample and $t_{0.5}$ is the half-rise time, the time required for the rear-surface temperature rise to reach one half of its maximum (steady-state) value $T_{\infty}$ (Figure \ref{fig:laser_flash}d). 

Since 1961, many papers have presented advances and extensions on \citeauthor{parker_1961}'s original method. In particular, numerous papers have focussed on the \textit{finite pulse-time effect}, that is, the observation that the heat pulse occurs over a finite duration \cite{heckman_1973,cape_1963} and is rarely well approximated by an ideal instantaneous pulse (as assumed by \citet{parker_1961}). In the one-dimensional case, the boundary condition imposing thermal insulation at the front surface \cite{parker_1961,carr_2019} is replaced by a boundary condition specifying the heat flux applied at the front surface based on the actual shape of the heat pulse  \cite{heckman_1973} (Figure \ref{fig:laser_flash}a). Commonly, simple functions are used to represent the pulse shape such as those describing rectangular \cite{azumi_1981}, triangular \cite{heckman_1973,azumi_1981} and exponential \cite{larson_1968} pulses. For the half-rise time approach, accounting for the finite-pulse time leads to small corrections \cite{azumi_1981,heckman_1973} to the formula for the thermal diffusivity (\ref{eq:alpha_parker}).

Considerable research activity has also focussed on the application of the laser flash method to layered samples consisting of two or more adjacent homogeneous layers with distinct thermal diffusivities \cite{larson_1968,james_1980,lee_1975,ohta_1990}. For a two layer sample, only one of the thermal diffusivities can be calculated from the temperature rise recorded at the rear surface \cite{koski_1985,lee_1975}. In this case, assuming the thermal diffusivity of one of the layers is known, extension of the half-rise time approach yields a formula for the thermal diffusivity of the other layer involving the half-rise time and the volumetric heat capacity of both layers \cite{larson_1968}.

Recently, \citet{carr_2019} presented a method for calculating the thermal diffusivity from the rear-surface temperature rise history. This so-called \textit{rear-surface integral method} expresses the thermal diffusivity exactly in terms of the area between the theoretical rear-surface temperature rise curve, $T(L,t)$, and the steady-state temperature, $T_{\infty}$, over time (see Figure \ref{fig:laser_flash}d). For a homogeneous and thermally insulated sample, where the heat pulse is instantaneously and uniformly absorbed by a thin layer of thickness $\ell$ at the front surface, the thermal diffusivity formula is given by \cite{carr_2019}:
\begin{gather}
\label{eq:alpha_carr}
\alpha = \frac{T_{\infty}\bigl(L^{2}-\ell^{2}\bigr)}{6\int_{0}^{\infty} \left[T_{\infty} - T(L,t)\right]\,\text{d}t}.
\end{gather}
Numerical experiments carried out on synthetic data generated by adding Gaussian noise to the theoretical rear-surface temperature rise curve, $T(L,t)$, demonstrated that the formula (\ref{eq:alpha_carr}) produces estimates of the thermal diffusivity that are more accurate than the standard formula (\ref{eq:alpha_parker}).

\revision{For the case of an ideal instantaneous heat pulse ($\ell = 0$), the formula (\ref{eq:alpha_carr}) reduces to one derived by \citet{baba_2009} using their \textit{areal heat diffusion method}. The areal heat diffusion method obtains a closed-form expression for the area $\int_{0}^{\infty} \left[T_{\infty} - T(L,t)\right]\,\text{d}t = \lim_{s\rightarrow 0}[s^{-1}T_{\infty} - \widetilde{T}_{r}(s)]$ involving $\widetilde{T}_{r}(s)$ the Laplace transform of $T_{r}(t) := T(L,t)$. Our rear-surface integral method for deriving (\ref{eq:alpha_carr}) avoids explicit calculation of $T_{r}(t)$ and its Laplace transform. This means that it is easier to apply to more sophisticated heat flow models as considered in the current paper.}

\citet{carr_2019} essentially provided a proof-of-concept for the rear-surface integral approach applied to the classical heat flow model of the laser flash experiment \cite{parker_1961}. In the current paper, we present two key contributions. Firstly, we extend the rear-surface integral approach to account for finite pulse time effects deriving new formulas for the thermal diffusivity for rectangular, triangular, exponential and arbitrary pulse shapes (Section \ref{sec:single_layer}). Secondly, we extend these new theoretical results to two layer samples, developing new formulas for the thermal diffusivity of both layers (Section \ref{sec:two_layer}). After deriving the new formulas, we verify their accuracy and demonstrate their application to the types of experimental data that arise from the laser flash experiment (Section \ref{sec:results}). We then conclude with a summary of the work and a discussion regarding the limitations of the new formulas (Section \ref{sec:conclusions}).

\section{Homogeneous sample}
\label{sec:single_layer}
\subsection{Heat flow model}
Consider a thermally insulated homogeneous sample of uniform thickness $L$, thermal conductivity $k$, volumetric heat capacity $\rho c$ and thermal diffusivity $\alpha = k/(\rho c)$ (Figure \ref{fig:laser_flash}b). Let $\mathcal{T}(x,t)$ be the temperature of the sample at location $0 \leq x \leq L$ and time $t \geq 0$ and assume the heat pulse applied at the front surface, $x = 0$, is initiated at $t = 0$. We assume the temperature rise, $T(x,t) = \mathcal{T}(x,t) - T_{0}$, above the initial uniform temperature of the sample, $T_{0}$, resulting from the heat pulse satisfies the following heat flow model \cite{heckman_1973,azumi_1981,czel_2013}:
\begin{gather}
\label{eq:model1_pde}
\frac{\partial T}{\partial t}(x,t) = \alpha\frac{\partial^{2}T}{\partial x^{2}}(x,t),\quad 0 < x < L,\quad t > 0,\\
\label{eq:model1_ic}
T(x,0) = 0,\quad 0 < x < L,\\
\label{eq:model1_bcs}
-k\frac{\partial T}{\partial x}(0,t) = q(t),\quad\frac{\partial T}{\partial x}(L,t) = 0,
\end{gather}
where $q(t)$ is the heat flux function representing the laser pulse applied at the front surface of the sample (Figure \ref{fig:laser_flash}a). Provided $\lim_{t\rightarrow\infty} q(t) = 0$ and the total amount of heat absorbed at the front surface
\begin{gather}
\label{eq:Q_inf}
Q_{\infty} = \int_{0}^{\infty}q(s)\,\text{d}s < \infty,
\end{gather}
the steady-state temperature rise is given by \cite{heckman_1973}:
\begin{gather}
\label{eq:model1_Tinf}
T_{\infty} = \lim_{t\rightarrow\infty}T(x,t) = \frac{Q_{\infty}}{\rho c L},\quad 0 < x < L.
\end{gather}
This result is derived by first applying the thermally insulated boundary conditions (\ref{eq:model1_bcs}) that arise in the limit $t\rightarrow\infty$ to the linear steady-state solution of the heat equation (\ref{eq:model1_pde}) yielding $\lim_{t\rightarrow\infty} T(x,t) = T_{\infty}$, where $T_{\infty}$ is a constant.
The value of $T_{\infty}$ is then identified by noting that the change in the amount of heat in the sample must be balanced by the amount of heat entering through the front surface:
\begin{gather}
\label{eq:model1_heat_conservation}
\rho c\int_{0}^{L} T(x,t)\,\text{d}x = Q(t),
\end{gather}
where 
\begin{gather}
\label{eq:Qt}
Q(t) = \int_{0}^{t} q(s)\,\text{d}s.
\end{gather}
At steady state, Eq (\ref{eq:model1_heat_conservation}) reduces to $\rho c L T_{\infty} = Q_{\infty}$ yielding the stated result (\ref{eq:model1_Tinf}).

\subsection{Thermal diffusivity}
\label{sec:single_layer_diffusivity}
We now extend the rear-surface integral method for calculating thermal diffusivity \citep{carr_2019} to non-instantaneous heat pulse durations. The aim of this method is to derive a formula for the thermal diffusivity, $\alpha$, in terms of an integral involving $T(L,t)$, the rear-surface temperature rise extracted from the heat flow model (\ref{eq:model1_pde})--(\ref{eq:model1_bcs}). The starting point for the derivation is the function:
\begin{gather}
\label{eq:u_definition}
u(x) = \int_{0}^{\infty} \left[T_{\infty} - T(x,t)\right]\,\text{d}t.
\end{gather}
Differentiating $u(x)$ twice with respect to $x$, noting $T_{\infty}$ (\ref{eq:model1_Tinf}) is a constant and using the heat equation (\ref{eq:model1_pde}) and initial condition (\ref{eq:model1_ic}) yields the differential equation \cite{carr_2019}:
\begin{gather}
\label{eq:model1_u_ode}
\frac{\text{d}^{2}u}{\text{d}x^{2}} = -\frac{T_{\infty}}{\alpha},\quad 0 < x < L,
\end{gather}
which identifies $u(x)$ as a quadratic function:
\begin{gather}
\label{eq:u_polynomial}
u(x) = c_{0} + c_{1}x - \frac{T_{\infty}x^{2}}{2\alpha}.
\end{gather}
To identify the integration constants, $c_{0}$ and $c_{1}$, we combine the form of $\text{d}u/\text{d}x$ with the boundary conditions (\ref{eq:model1_bcs}) \cite{carr_2019} yielding:
\begin{align}
\label{eq:model1_u_bc1}
\frac{\text{d}u}{\text{d}x}(0) &= \int_{0}^{\infty} -\frac{\partial T}{\partial x}(0,t)\,\text{d}t = \int_{0}^{\infty} \frac{q(t)}{k}\,\text{d}t = \frac{Q_{\infty}}{k},\\
\label{eq:model1_u_bc2}
\frac{\text{d}u}{\text{d}x}(L) &= \int_{0}^{\infty} -\frac{\partial T}{\partial x}(L,t)\,\text{d}t = \int_{0}^{\infty} 0\,\text{d}t = 0.
\end{align}
Imposing both of these boundary conditions on $u(x)$ (\ref{eq:u_polynomial}) yields the same result, namely $c_{1} = T_{\infty}L/\alpha$, after noting the relationship between $T_{\infty}$ and $Q_{\infty}$ (\ref{eq:model1_Tinf}). Therefore, to uniquely identify $u(x)$ we require an auxiliary condition analogous to the one required for the steady-state solution (\ref{eq:model1_heat_conservation}). Integrating $u(x)$ (\ref{eq:u_definition}) from $x = 0$ to $x = L$, multiplying by the volumetric heat capacity $\rho c$ and reversing the order of integration yields the appropriate condition:
\begin{align}
\label{eq:model1_u_ac}
\rho c\int_{0}^{L} u(x)\,\text{d}x = \int_{0}^{\infty} \left[Q_{\infty} - Q(t)\right]\,\text{d}t.
\end{align}
Imposing this auxiliary condition on $u(x)$ (\ref{eq:u_polynomial}) uniquely identifies $c_{0}$ and hence $u(x)$:  
\begin{gather}
\label{eq:model1_u_exp2}
u(x) = \frac{\int_{0}^{\infty}[Q_{\infty} - Q(t)]\,\text{d}t}{\rho c L} + \frac{T_{\infty}L}{\alpha}\left[x - \frac{L}{3} - \frac{x^{2}}{2L}\right].
\end{gather}
In summary, we have derived a closed-form expression for the integral (\ref{eq:u_definition}) by formulating and solving the boundary value problem consisting of Eqs (\ref{eq:model1_u_ode}) and (\ref{eq:model1_u_bc1})--(\ref{eq:model1_u_ac}). Equating both expressions for $u(x)$, Eqs (\ref{eq:u_definition}) and (\ref{eq:model1_u_exp2}), at the rear surface, $x = L$, and rearranging yields a formula for the thermal diffusivity that can be expressed as:
\begin{gather}
\label{eq:model1_alpha}
\alpha = \frac{L^{2}}{6(I_{T}-I_{q})},
\end{gather}
where
\begin{gather}
\label{eq:model1_IT}
I_{T} = \int_{0}^{\infty}[1 - T(L,t)/T_{\infty}]\,\text{d}t,\\
\label{eq:model1_Iq}
I_{q} = \int_{0}^{\infty}[1 - Q(t)/Q_{\infty}]\,\text{d}t.
\end{gather}
Eqs (\ref{eq:model1_alpha})--(\ref{eq:model1_Iq}) express the thermal diffusivity explicitly in terms of $L$, the thickness of the sample, $T(L,t)$, the rear-surface temperature rise history, $T_{\infty}$, the steady-state temperature rise value (\ref{eq:model1_Tinf}), $Q_{\infty}$, the total amount of heat absorbed through the front surface (\ref{eq:Q_inf}), and $Q(t)$, the amount of heat that has been absorbed through the front surface at time $t$ (\ref{eq:Qt}). For the types of functions commonly used to represent the shape of the heat pulse, which specify the finite pulse time $\tau$, and/or peak of the pulse $\beta$ (Figure \ref{fig:laser_flash}a), the formula for the thermal diffusivity (\ref{eq:model1_alpha})--(\ref{eq:model1_Iq}) simplifies significantly:
\medskip

\noindent\textit{Rectangular pulse} \citep{azumi_1981}:
\begin{subequations}
\label{eq:rectangular_pulse}
\begin{align}
q(t) &= \begin{cases} \dfrac{Q_{\infty}}{\tau}, & 0 < t \leq \tau,\\ 0, & \text{otherwise}, \end{cases}\qquad I_{q} = \frac{\tau}{2},\\
\alpha &= \frac{L^{2}}{6(I_{T}-\frac{\tau}{2})}.
\end{align}
\end{subequations}
\textit{Triangular pulse} \citep{azumi_1981,heckman_1973}:
\begin{subequations}
\label{eq:triangular_pulse}
\begin{align}
q(t) &= \begin{cases} \dfrac{2Q_{\infty}t}{\tau\beta}, & 0 < t \leq \beta,\\[0.25cm] \dfrac{2Q_{\infty}(\tau-t)}{\tau(\tau-\beta)}, & \beta\leq t\leq\tau,\\[0.25cm] 0, & \text{otherwise},\end{cases}\quad I_{q} = \frac{\tau+\beta}{3},\\
\alpha &= \frac{L^{2}}{6\bigl(I_{T} - \frac{\tau+\beta}{3}\bigr)}.
\end{align}
\end{subequations}
\textit{Exponential pulse} \citep{larson_1968}:
\begin{subequations}
\label{eq:exponential_pulse}
\begin{align}
q(t) &= \begin{cases} \dfrac{Q_{\infty}t}{\beta^{2}}e^{-t/\beta}, & t > 0,\\ 0, & \text{otherwise}, \end{cases}\quad I_{q} = 2\beta,\\
\alpha &= \frac{L^{2}}{6(I_{T} - 2\beta)}.
\end{align}
\end{subequations}
In the limit $\tau\rightarrow 0$, the rectangular pulse tends to the instantaneous pulse $q(t) = Q_{\infty}\delta(t)$, where $\delta(t)$ is the Dirac delta function, yielding $\alpha = L^{2}/(6I_{T})$ as derived previously \cite{carr_2019}. Hence, the above formulas provide simple finite pulse time corrections to the thermal diffusivity involving the duration and peak of the pulse. Application of the thermal diffusivity formula (\ref{eq:model1_alpha})--(\ref{eq:model1_Iq}) to arbitrary pulse shapes is addressed later in Section \ref{sec:noisy_data}.

\revision{An interesting question is the following: if one assumes the heat pulse occurs instantaneously but actually occurs over a finite duration, what is the effect on the error in the thermal diffusivity estimate? The answer to this question is evident from the formulas for $\alpha$ in Eqs (\ref{eq:rectangular_pulse})--(\ref{eq:exponential_pulse}). For example, if the true pulse was triangular, the thermal diffusivity would be estimated by $\alpha_{I} = L^{2}/(6I_{T})$ instead of $\alpha_{T} = L^{2}/[6(I_{T} - \frac{\tau+\beta}{3})]$ leading to an underestimate since $\tau$ and $\beta$ are both positive numbers. Going further, the relative error in the calculation can be expressed as $\left|\alpha_{I} - \alpha_{T}\right|/\alpha_{T} = 2(\tau+\beta)\alpha_{I}/L^{2}$. For the parameter values considered in this paper (Table \ref{tab:parameter_values}), this result leads to a large relative error of approximately $27 \%$, which highlights the importance of incorporating finite pulse time effects in the calculations.} 

\section{Two layer sample}
\label{sec:two_layer}
 
\subsection{Heat flow model}
We now consider a two-layer sample comprised of two homogeneous materials with constant but different thermodynamic properties. The first layer has thickness $\ell_{1}$ and extends from the front surface of the sample at $x = 0$ to the interface between the two layers at $x = \ell_{1}$ while the second layer has thickness $\ell_{2}$ and extends from the interface at $x = \ell_{1}$ to the rear surface of the sample at $x = \ell_{1} + \ell_{2} =: L$ (Figure \ref{fig:laser_flash}c). The first layer is assumed to have thermal conductivity $k_{1}$, volumetric heat capacity $\rho_{1}c_{1}$ and thermal diffusivity $\alpha_{1} = k_{1}/(\rho_{1}c_{1})$ while the second layer has thermal conductivity $k_{2}$, volumetric heat capacity $\rho_{2}c_{2}$ and thermal diffusivity $\alpha_{2} = k_{2}/(\rho_{2}c_{2})$. Let $\mathcal{T}_{1}(x,t)$ be the temperature of the sample in the first layer at location $0\leq x\leq \ell_{1}$ and time $t \geq 0$ and $\mathcal{T}_{2}(x,t)$ be the temperature of the sample in the second layer at location $\ell_{1}\leq x\leq L$ and time $t \geq 0$. We assume the temperature rise in the first and second layers, $T_{1}(x,t) = \mathcal{T}_{1}(x,t) - T_{0}$ and $T_{2}(x,t) = \mathcal{T}_{2}(x,t) - T_{0}$, after initiation of the heat pulse at $t = 0$, satisfies the following heat flow model \cite{larson_1968,lee_1975}: 
\begin{gather}
\label{eq:model2_pde1}
\frac{\partial T_{1}}{\partial t}(x,t) = \alpha_{1}\frac{\partial^{2}T_{1}}{\partial x^{2}}(x,t),\quad 0 < x < \ell_{1},\quad t > 0,\\
\label{eq:model2_pde2}
\frac{\partial T_{2}}{\partial t}(x,t) = \alpha_{2}\frac{\partial^{2}T_{2}}{\partial x^{2}}(x,t),\quad \ell_{1} < x < L,\quad t > 0,\\
\label{eq:model2_ic1}
T_{1}(x,0) = 0,\quad 0 < x < \ell_{1},\\
\label{eq:model2_ic2}
T_{2}(x,0) = 0,\quad \ell_{1} < x < L,\\
\label{eq:model2_bc1}
-k_{1}\frac{\partial T_{1}}{\partial x}(0,t) = q(t),\quad t>0,\\
\label{eq:model2_bc2}
\frac{\partial T_{2}}{\partial x}(L,t) = 0,\quad t>0,\\
\label{eq:model2_intc1}
T_{1}(\ell_{1},t) = T_{2}(\ell_{1},t),\quad t > 0,\\
\label{eq:model2_intc2}
k_{1}\frac{\partial T_{1}}{\partial x}(\ell_{1},t) = k_{2}\frac{\partial T_{2}}{\partial x}(\ell_{1},t),\quad t > 0,
\end{gather}
where the internal boundary conditions (\ref{eq:model2_intc1})--(\ref{eq:model2_intc2}) specify continuity of temperature and heat flux at the interface between the two layers.

In a similar manner to the homogeneous case, provided $\lim_{t\rightarrow\infty} q(t) = 0$ and the total amount of heat $Q_{\infty}$ absorbed at the front surface (\ref{eq:Q_inf}) is finite, the steady-state solution of the two-layer heat flow model (\ref{eq:model2_pde1})--(\ref{eq:model2_intc2}) is given by \cite{larson_1968,chen_2010}:
\begin{gather}
\label{eq:model2_Tinf}
T_{\infty} = \lim_{t\rightarrow\infty} T_{1}(x,t) = \lim_{t\rightarrow\infty} T_{2}(x,t) = \frac{Q_{\infty}}{\rho_{1}c_{1}\ell_{1} + \rho_{2}c_{2}\ell_{2}}.
\end{gather}
This result is derived by applying the interface conditions (\ref{eq:model2_intc1})--(\ref{eq:model2_intc2}) and the thermally insulated boundary conditions (\ref{eq:model2_bc1})--(\ref{eq:model2_bc2}) that arise in the limit $t\rightarrow\infty$ to the linear steady-state solutions of the heat equations (\ref{eq:model2_pde1})--(\ref{eq:model2_pde2}) yielding $\lim_{t\rightarrow\infty} T_{1}(x,t) = \lim_{t\rightarrow\infty} T_{2}(x,t) = T_{\infty}$, where $T_{\infty}$ is a constant. As with the homogeneous case, the value of $T_{\infty}$ can be identified by noting that the change in the amount of heat in the sample must be balanced by the amount of heat entering the sample through the front surface:
\begin{gather}
\label{eq:model2_heat_conservation}
\rho_{1}c_{1}\int_{0}^{\ell_{1}} T_{1}(x,t)\,\text{d}x + \rho_{2}c_{2}\int_{\ell_{1}}^{L} T_{2}(x,t)\,\text{d}x = Q(t),
\end{gather}
where $Q(t)$ is as defined previously (\ref{eq:Qt}). At steady state this heat conservation statement reduces to $T_{\infty}\left[\rho_{1}c_{1}\ell_{1} + \rho_{2}c_{2}\ell_{2}\right] = Q_{\infty}$ yielding the stated result (\ref{eq:model2_Tinf}).

\subsection{Thermal diffusivity}
\label{sec:two_layer_diffusivity}
We now extend the analysis of Section \ref{sec:single_layer_diffusivity} to two-layer samples. The aim is again to derive a formula for the thermal diffusivity in terms of an integral involving the theoretical rear-surface temperature rise history, which for the two-layer heat flow model (\ref{eq:model2_pde1})--(\ref{eq:model2_intc2}) is represented by $T_{2}(L,t)$. For the two-layer case, we have separate definitions of Eq (\ref{eq:u_definition}) in each layer:
\begin{align}
\label{eq:u1_definition}
u_{1}(x) = \int_{0}^{\infty}\left[T_{\infty} - T_{1}(x,t)\right]\,\text{d}t,\\
\label{eq:u2_definition}
u_{2}(x) = \int_{0}^{\infty}\left[T_{\infty} - T_{2}(x,t)\right]\,\text{d}t.
\end{align}
Using an identical procedure to the one carried out for the single-layer case yields the following differential equations and boundary conditions:
\begin{gather}
\label{eq:model2_u_de1}
\frac{\text{d}^{2}u_{1}}{\text{d}x^{2}} = -\frac{T_{\infty}}{\alpha_{1}},\quad 0 < x < \ell_{1},\\
\label{eq:model2_u_de2}
\frac{\text{d}^{2}u_{2}}{\text{d}x^{2}} = -\frac{T_{\infty}}{\alpha_{2}},\quad \ell_{1} < x < L,\\
\label{eq:model2_u_bcs}
\frac{\text{d}u_{1}}{\text{d}x}(0) = \frac{Q_{\infty}}{k_{1}},\quad \frac{\text{d}u_{2}}{\text{d}x}(L) = 0.
\end{gather}
The solution of Eqs (\ref{eq:model2_u_de1})--(\ref{eq:model2_u_bcs}) is
\begin{align}
\label{eq:u1_intermediate}
u_{1}(x) &= a_{0} + \frac{T_{\infty}(\rho_{1}c_{1}\ell_{1} + \rho_{2}c_{2}\ell_{2})x}{\alpha_{1}\rho_{1}c_{1}} - \frac{T_{\infty}x^{2}}{2\alpha_{1}},\\
\label{eq:u2_intermediate}
u_{2}(x) &= b_{0} + \frac{T_{\infty}(\ell_{1}+\ell_{2})x}{\alpha_{2}} - \frac{T_{\infty}x^{2}}{2\alpha_{2}},
\end{align}
where $a_{0}$ and $b_{0}$ are arbitrary constants. The appropriate interface conditions are given by
\begin{gather}
\label{eq:model2_u_ics1}
u_{1}(\ell_{1}) = u_{2}(\ell_{1}),\\
\label{eq:model2_u_ics2}
k_{1}\frac{\text{d}u_{1}}{\text{d}x}(\ell_{1}) = k_{2}\frac{\text{d}u_{2}}{\text{d}x}(\ell_{1}),
\end{gather}
which are derived by combining the interface conditions of the heat flow model (\ref{eq:model2_intc1})--(\ref{eq:model2_intc2}) with the definitions of $u_{1}(x)$ and $u_{2}(x)$ (\ref{eq:u1_definition})--(\ref{eq:u2_definition}). For example, the second interface condition (\ref{eq:model2_u_ics2}) is derived as follows:
\begin{align*}
k_{1}\frac{\text{d}u_{1}}{\text{d}x}(\ell_{1}) &= \int_{0}^{\infty} -k_{1}\frac{\partial T_{1}}{\partial x}(\ell_{1},t)\,\text{d}t\\ 
&= \int_{0}^{\infty} -k_{2}\frac{\partial T_{2}}{\partial x}(\ell_{1},t)\,\text{d}t = k_{2}\frac{\text{d}u_{2}}{\text{d}x}(\ell_{1}).
\end{align*}
Both $u_{1}(x)$ and $u_{2}(x)$ (\ref{eq:u1_intermediate})--(\ref{eq:u2_intermediate}) already satisfy flux continuity at the interface (\ref{eq:model2_u_ics2}) and therefore this interface condition yields no information about $a_{0}$ and $b_{0}$. As with the single-layer case, an additional condition is required to completely identify the solution. The analogue of the auxiliary condition (\ref{eq:model1_u_ac}) for the two-layer case is:
\begin{multline}
\label{eq:model2_u_ac}
\rho_{1}c_{1}\int_{0}^{\ell_{1}}u_{1}(x)\,\text{d}x + \rho_{2}c_{2}\int_{\ell_{1}}^{L}u_{2}(x)\,\text{d}x\\ = \int_{0}^{\infty}\left[Q_{\infty} - Q(t)\right]\,\text{d}t,
\end{multline}
which is derived by substituting the definitions of $u_{1}(x)$ and $u_{2}(x)$ (\ref{eq:u1_definition})--(\ref{eq:u2_definition}) into the left-hand side of Eq (\ref{eq:model2_u_ac}), reversing the order of integration, combining the integrals and making use of the expression for $T_{\infty}$ (\ref{eq:model2_Tinf}) and the heat conservation statement (\ref{eq:model2_heat_conservation}). Applying the auxiliary equation (\ref{eq:model2_u_ac}) and the first interface condition (\ref{eq:model2_u_ics1}), produces a pair of linear equations that allow $a_{0}$ and $b_{0}$ to be identified. This determines the final forms for $u_{1}(x)$ and $u_{2}(x)$, which together comprise the solution to the boundary value problem consisting of Eqs (\ref{eq:model2_u_de1})--(\ref{eq:model2_u_bcs}) and (\ref{eq:model2_u_ics1})--(\ref{eq:model2_u_ac}). Since we are only interested in the rear-surface behaviour, we give only the final expression for $u_{2}(x)$: 
\begin{align}
\nonumber
u_{2}(x) &= \frac{\int_{0}^{\infty}[Q_{\infty} - Q(t)]\,\text{d}t}{\rho_{1}c_{1}\ell_{1} + \rho_{2}c_{2}\ell_{2}} - \frac{\ell_{1}^{2}[2T_{\infty}\rho_{1}c_{1}\ell_{1} - 3Q_{\infty}]}{6\alpha_{1}(\rho_{1}c_{1}\ell_{1}+\rho_{2}c_{2}\ell_{2})}\\ \nonumber &\quad - \frac{\rho_{1}c_{1}\ell_{1}^{2}T_{\infty}\left[\ell_{1} + 2\ell_{2}\right]}{2\alpha_{2}\left(\rho_{1}c_{1}\ell_{1} + \rho_{2}c_{2}\ell_{2}\right)}\\ 
\nonumber &\quad - \frac{\rho_{2}c_{2}T_{\infty}[2(\ell_{1}+\ell_{2})^{3} - \ell_{1}^{2}(2\ell_{1}+3\ell_{2})]}{6\alpha_{2}(\rho_{1}c_{1}\ell_{1}+\rho_{2}c_{2}\ell_{2})}\\ \label{eq:u2_final} &\quad+ \frac{T_{\infty}(\ell_{1}+\ell_{2})x}{\alpha_{2}} - \frac{T_{\infty}x^{2}}{2\alpha_{2}}.
\end{align}
Equating both expressions for $u_{2}(x)$, Eqs (\ref{eq:u2_definition}) and (\ref{eq:u2_final}), at the rear surface, $x = L$, and rearranging yields formulas for the thermal diffusivity in either the first or second layers:
\begin{align}
\label{eq:model2_alpha1}
\alpha_{1} &= \frac{\ell_{1}^{2}(\ell_{1}\rho_{1}c_{1} + 3\ell_{2}\rho_{2}c_{2})\alpha_{2}}{6\alpha_{2}(\rho_{1}c_{1}\ell_{1}+\rho_{2}c_{2}\ell_{2})\left(I_{T} - I_{q}\right) - \ell_{2}^{2}(3\ell_{1}\rho_{1}c_{1}+\ell_{2}\rho_{2}c_{2})},\\
\label{eq:model2_alpha2}
\alpha_{2} &= \frac{\ell_{2}^{2}(3\ell_{1}\rho_{1}c_{1}+\ell_{2}\rho_{2}c_{2})\alpha_{1}}{6\alpha_{1}(\rho_{1}c_{1}\ell_{1}+\rho_{2}c_{2}\ell_{2})(I_{T}-I_{q}) - \ell_{1}^{2}[\ell_{1}\rho_{1}c_{1}+3\ell_{2}\rho_{2}c_{2}]},
\end{align}
where 
\begin{gather}
\label{eq:model2_IT}
I_{T} = \int_{0}^{\infty}[1-T_{2}(L,t)/T_{\infty}]\,\text{d}t,\\
\label{eq:model2_Iq}
\text{$I_{q}$ is as defined in Eq (\ref{eq:model1_Iq})}.
\end{gather}
Both formulas (\ref{eq:model2_alpha1})--(\ref{eq:model2_alpha2}) assume the layer thicknesses, $\ell_{1}$ and $\ell_{2}$, and the volumetric heat capacities, $\rho_{1}c_{1}$ and $\rho_{2}c_{2}$, are known, expressing the thermal diffusivity explicitly in terms of $T_{2}(L,t)$, the rear-surface temperature rise history, $T_{\infty}$ the steady-state temperature rise value (\ref{eq:model2_Tinf}), $Q_{\infty}$ the total amount of heat absorbed through the front surface (\ref{eq:Q_inf}), and $Q(t)$ the amount of heat that has been absorbed through the front surface at time $t$ (\ref{eq:Qt}). Calculating the thermal diffusivity of either layer also requires the thermal diffusivity of the other layer to be known, which is consistent with the well-known result that only one thermal diffusivity of a two-layer sample can be calculated using the rear-surface temperature \cite{koski_1985,lee_1975}. Since the formula for $I_{q}$ is identical for the homogeneous and two-layer case, the two-layer thermal diffusivity formulas for the rectangular, triangular and exponential pulses, analogous to Eqs (\ref{eq:rectangular_pulse})--(\ref{eq:exponential_pulse}), are easily obtained by substituting the expressions for $I_{q}$ from Eqs (\ref{eq:rectangular_pulse})--(\ref{eq:exponential_pulse}) into Eqs (\ref{eq:model2_alpha1})--(\ref{eq:model2_alpha2}). Finally, we confirm that removing the heterogeneity by setting $\alpha_{1} = \alpha_{2} = \alpha$, $\rho_{1}c_{1} = \rho_{2}c_{2} = \rho c$ and $T_{2}(L,t) = T(L,t)$ in both formulas (\ref{eq:model2_alpha1})--(\ref{eq:model2_alpha2}) and rearranging for $\alpha$ yields the single-layer formula (\ref{eq:model1_alpha}) as expected.

\section{Numerical experiments}
\label{sec:results}

\subsection{Verification of formulas}
\label{sec:verification}
We now verify the thermal diffusivity formulas derived in Sections \ref{sec:single_layer_diffusivity} and \ref{sec:two_layer_diffusivity} for the homogeneous and two-layer samples. In practice, the rear-surface temperature rise history takes the form of discrete values: $\widetilde{T}_{0}, \hdots, \widetilde{T}_{N}$, where $\widetilde{T}_{i}$ is the sampled value at $t_{i} = i\Delta t$, $\Delta t = t_{N}/N$ and $t_{N}$ is the final sampled time. In the discrete case, provided $t_{N}$ is large enough (see, e.g. \cite{carr_2017,carr_2018_joh}, for how to estimate steady state times), the integral $I_{T}$ appearing in the thermal diffusivity formulas, Eqs (\ref{eq:model1_alpha})--(\ref{eq:model1_Iq}) and (\ref{eq:model2_alpha1})--(\ref{eq:model2_Iq}), can be approximated using the trapezoidal rule \cite{carr_2019}. For example, for the homogeneous sample:
\begin{align}
\nonumber
I_{T} &\approx \int_{0}^{t_{N}}\left[1 - T(L,t)/T_{\infty}\right]\,\text{d}t,\\ 
\label{eq:IT_approx}
&\approx \Delta t\sum_{i=1}^{N} \left(1 - \frac{\widetilde{T}_{i-1}+\widetilde{T}_{i}}{2T_{\infty}}\right).
\end{align}
Verification is carried out using synthetic rear-surface temperature rise data generated from solving the heat flow model using a known set of parameter values (Table \ref{tab:parameter_values}) given previously by \citet{czel_2013}. Both the homogeneous model (\ref{eq:model1_pde})--(\ref{eq:model1_bcs}) and two-layer model (\ref{eq:model2_pde1})--(\ref{eq:model2_intc2}) are solved numerically using finite volume schemes consisting of $n$ nodes, as outlined in \ref{app:single_layer} and \ref{app:two_layer}, respectively. These numerical solutions produce sampled values of the rear-surface temperature rise at equally-spaced discrete times such that the $i$th sampled value $\widetilde{T}_{i}$ is given by $T_{n}(t_{i})$, where $T_{n}$ represents the temperature rise value in the finite volume scheme at the node located at the rear surface ($x = L$, $n$th node). Each estimate of the thermal diffusivity is compared to the target value in Table \ref{tab:parameter_values} by calculating the signed relative errors $\varepsilon = (\alpha - \widetilde{\alpha})/\alpha$ (homogeneous) and $\varepsilon_{k} = (\alpha_{k} - \widetilde{\alpha}_{k})/\alpha_{k}$ for $k = 1,2$ (two-layer) with the tilde used to indicate the estimated value produced from the rear-surface integral method. 

\begin{table*}[t]
\centering
\begin{tabular}{|c|}
\hline
\begin{minipage}{0.9\textwidth}
\vspace*{-0.3cm}
\begin{gather*}
\textit{Homogeneous (Single-layer)}\\
k = 222\,\text{W}\,\text{m}^{-1}\text{K}^{-1},\quad
\rho = 2700\,\text{kg}\,\text{m}^{-3},\quad c = 896\,\text{J}\,\text{kg}^{-1}\text{K}^{-1},\quad L = 0.002\,\text{m},\\
\text{Target value: $\alpha = k/(\rho c) = 9.1766\times 10^{-5}\,\text{m}^{2}\text{s}^{-1}$.}\\
\textit{Heterogeneous (Two-layer)}\\
k_{1} = 222\,\text{W}\,\text{m}^{-1}\text{K}^{-1},\quad \rho_{1} = 2700\,\text{kg}\,\text{m}^{-3},\quad c_{1} = 896\,\text{J}\,\text{kg}^{-1}\text{K}^{-1},\quad \ell_{1} = 0.00176\,\text{m},\\
k_{2} = 16.3\,\text{W}\,\text{m}^{-1}\text{K}^{-1},\quad \rho_{2} = 7810\,\text{kg}\,\text{m}^{-3},\quad c_{2} = 480\,\text{J}\,\text{kg}^{-1}\text{K}^{-1},\quad \ell_{2} = 0.00024\,\text{m},\\
\text{Target value: $\alpha_{1} = k_{1}/(\rho_{1}c_{1}) = 9.1766\times 10^{-5}\,\text{m}^{2}\text{s}^{-1}$,}\\
\text{Target value: $\alpha_{2} = k_{2}/(\rho_{2}c_{2}) = 4.3481\times 10^{-6}\,\text{m}^{2}\text{s}^{-1}$.}\\[-0.35cm]
\end{gather*}
\end{minipage}\\
\hline
\end{tabular}
\caption{Heat flow parameter values used in all numerical experiments. In each case, the target value of the thermal diffusivity is displayed rounded to five significant figures.}
\label{tab:parameter_values}
\end{table*}

\begin{table*}[h]
\centering
\renewcommand{\arraystretch}{1.0}
\begin{tabular*}{\textwidth}{@{\extracolsep{\fill}}lrrr}
\hline
 & Rectangular & Triangular & Exponential\\
\hline
Homogeneous (single-layer)\\
$\widetilde{\alpha}$ \revision{[$\text{m}^{2}\,\text{s}^{-1}$]} & \num{9.1766e-05} & \num{9.1766e-05} & \num{9.1766e-05}\\
$\varepsilon$ \revision{[\%]} & \num{2.0077e-04} & \num{2.0078e-04} & \num{2.0078e-04}\\
\hline
Heterogeneous (two layer)\\
$\widetilde{\alpha}_{1}$ \revision{[$\text{m}^{2}\,\text{s}^{-1}$]} & \num{9.1766e-05} & \num{9.1766e-05} & \num{9.1766e-05}\\
$\varepsilon_{1}$ \revision{[\%]} & \num{1.5619e-04} & \num{1.8399e-04} & \num{1.8065e-04}\\
$\widetilde{\alpha}_{2}$ \revision{[$\text{m}^{2}\,\text{s}^{-1}$]} & \num{4.3480e-06} & \num{4.3480e-06} & \num{4.3480e-06}\\
$\varepsilon_{2}$ \revision{[\%]} & \num{2.0250e-04} & \num{2.3855e-04} & \num{2.3421e-04}\\
\hline
\end{tabular*}
\caption{Thermal diffusivity estimates, $\widetilde{\alpha}$, $\widetilde{\alpha}_{1}$ and $\widetilde{\alpha}_{2}$, and corresponding signed relative errors, $\varepsilon$, $\varepsilon_{1}$ and $\varepsilon_{2}$, obtained from applying the formulas (\ref{eq:model1_alpha})--(\ref{eq:model1_Iq}) and (\ref{eq:model2_alpha1})--(\ref{eq:model2_Iq}) to the parameter values in Table \ref{tab:parameter_values} and the synthetic rear-surface temperature rise history discussed in Section \ref{sec:verification}. Results are given for the rectangular, triangular and exponential pulses given in Eqs (\ref{eq:rectangular_pulse})--(\ref{eq:exponential_pulse}).}
\label{tab:results1}
\end{table*}

In Table \ref{tab:results1}, results for both the homogeneous and two-layer samples are presented for the rectangular (\ref{eq:rectangular_pulse}), triangular (\ref{eq:triangular_pulse}) and exponential (\ref{eq:exponential_pulse}) pulses with $\tau = 0.005\,\text{s}$, $\beta = 0.001\,\text{s}$, $Q_{\infty} = 7000\,\text{J}\,\text{m}^{-2}$, $N = 1000$ temperature rise values, $\Delta t = 10^{-4}\,\text{s}$, $t_{N} = 0.1\,\text{s}$ and $n = 500$ nodes. In all cases, the estimate agrees with the target values of the thermal diffusivity in Table \ref{tab:parameter_values} to 4--5 significant digits with a relative error of approximately $10^{-4}$. These small non-zero values are explained by the error present in the numerical experiment incurred from truncating the infinite upper limit of the integral $I_{T}$ at the finite value $t_{N}$ and approximating the resulting integral numerically (\ref{eq:IT_approx}) as well as numerically solving the heat flow models to generate the sampled values. In summary, the results in Table \ref{tab:results1} confirm the correctness of the analysis in Sections \ref{sec:single_layer_diffusivity} and \ref{sec:two_layer_diffusivity}.

\subsection{Application to noisy data}
\label{sec:noisy_data}
We now investigate the accuracy of the thermal diffusivity formulas when applied to noisy rear-surface temperature data. To mimic the data arising from the laser flash experiment, we follow \citet{carr_2019} and add Gaussian noise to the sampled values from the previous section: $\widetilde{T}_{i} = T_{n}(t_{i}) + z_{i}$, where $z_{i}$ is a random number generated from a normal distribution with mean zero and standard deviation $\sigma$. Repeatedly generating synthetic rear-surface temperature rise datasets in this manner and calculating the corresponding thermal diffusivities allows distributions of the signed relative error to be constructed. 

\begin{figure*}[t]
\centering
\subfloat[]{\includegraphics[width=0.31\textwidth]{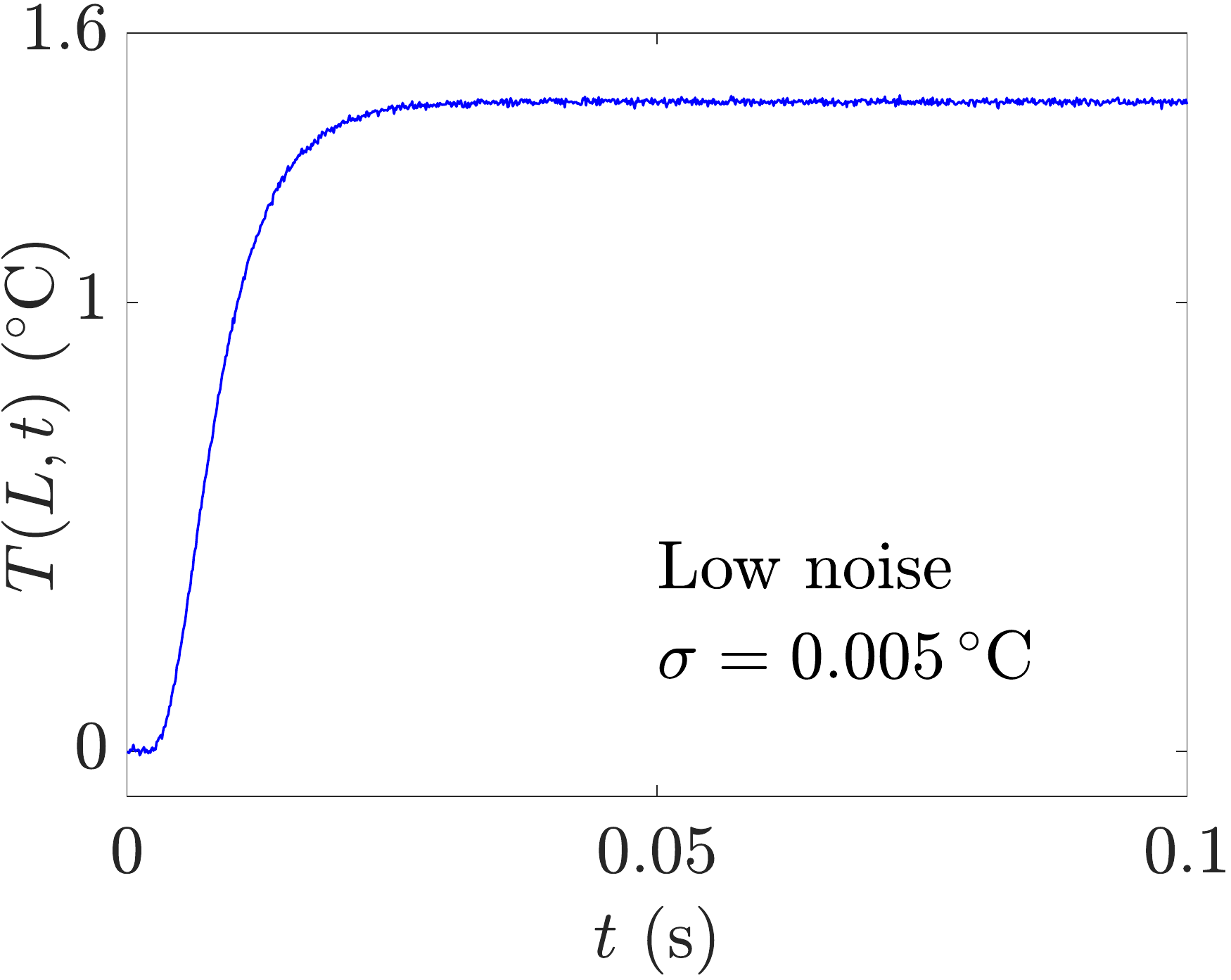}}\hfill\subfloat[]{\includegraphics[width=0.31\textwidth]{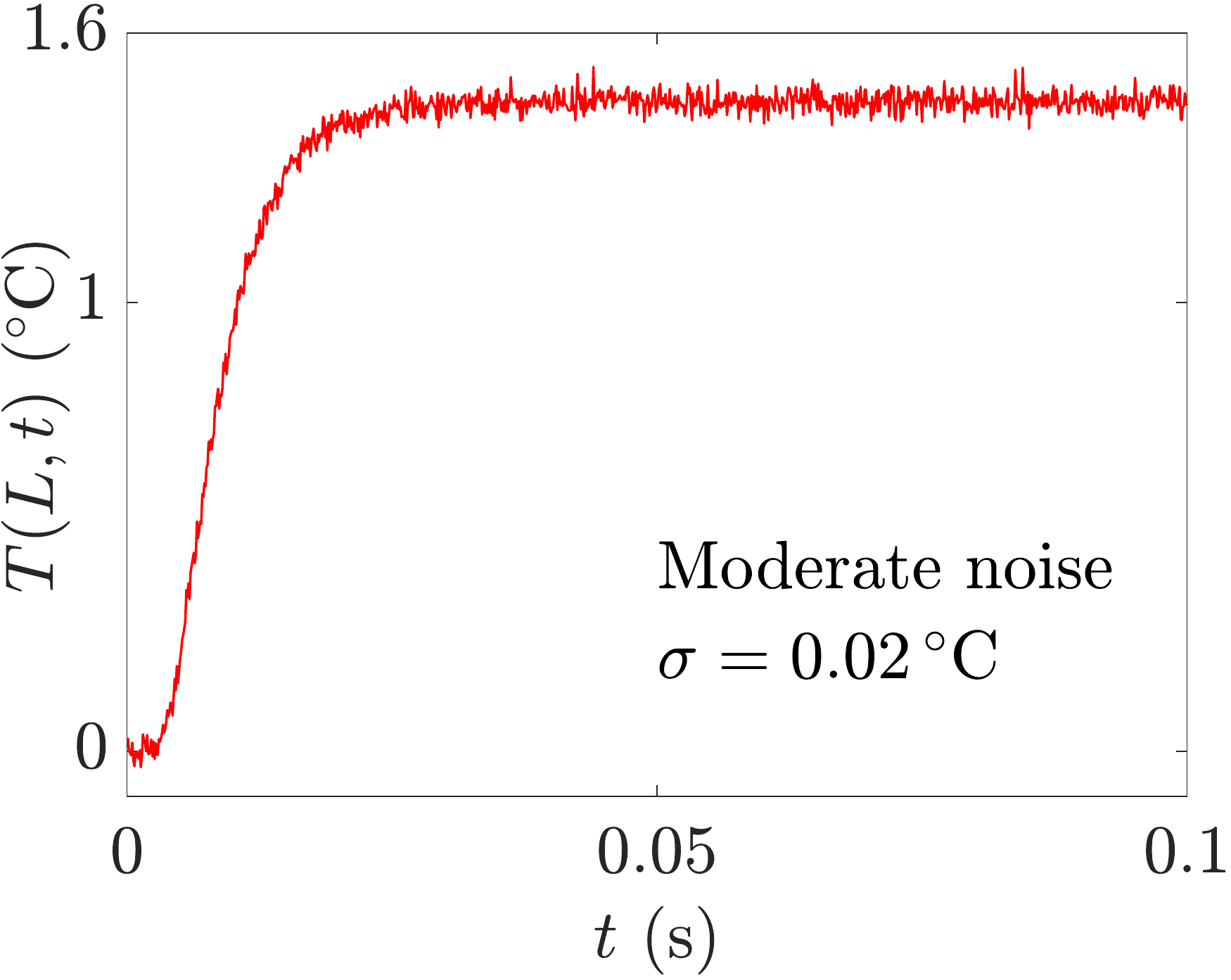}}\hfill\subfloat[]{\includegraphics[width=0.31\textwidth]{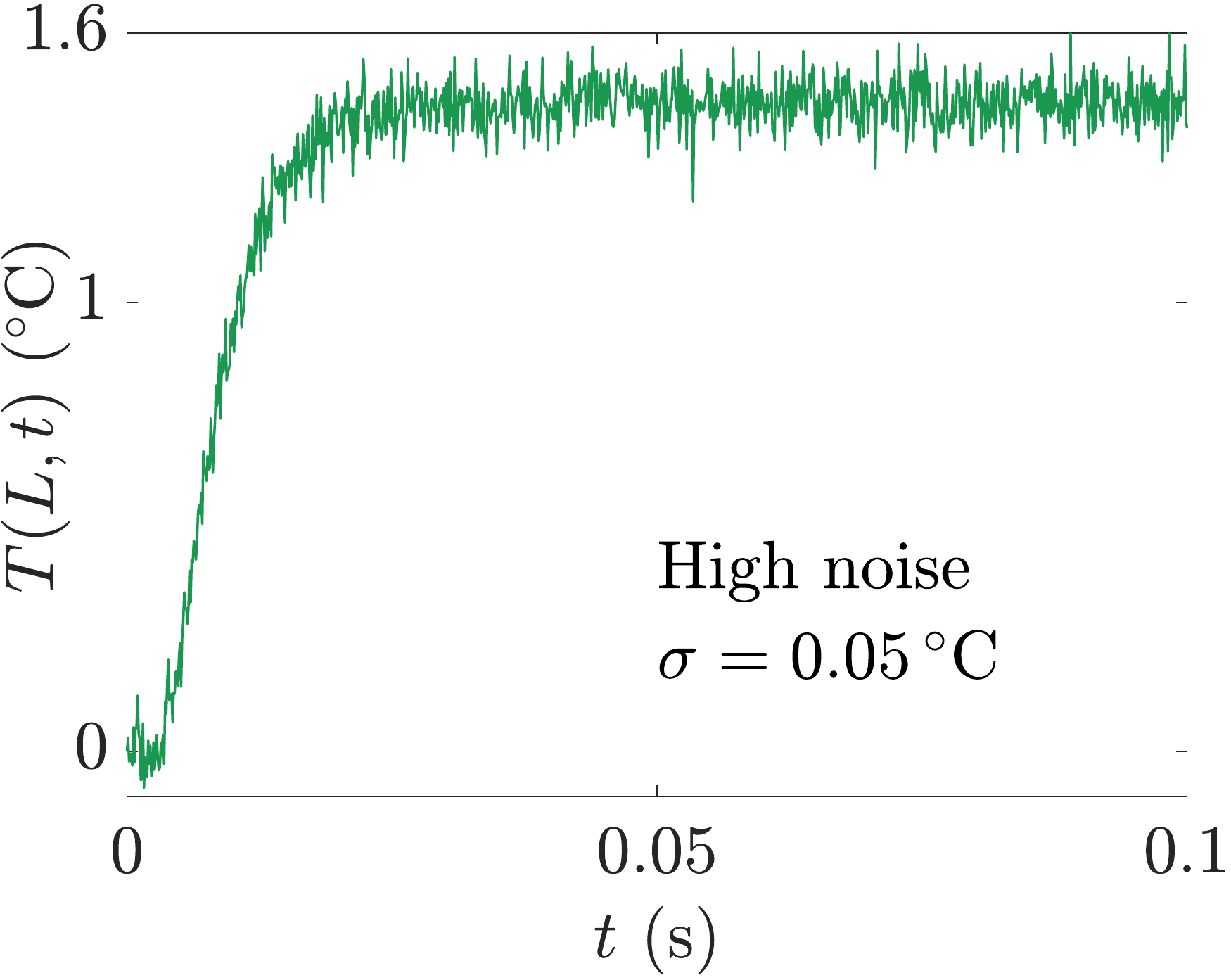}}\\
\subfloat[]{\includegraphics[width=0.31\textwidth]{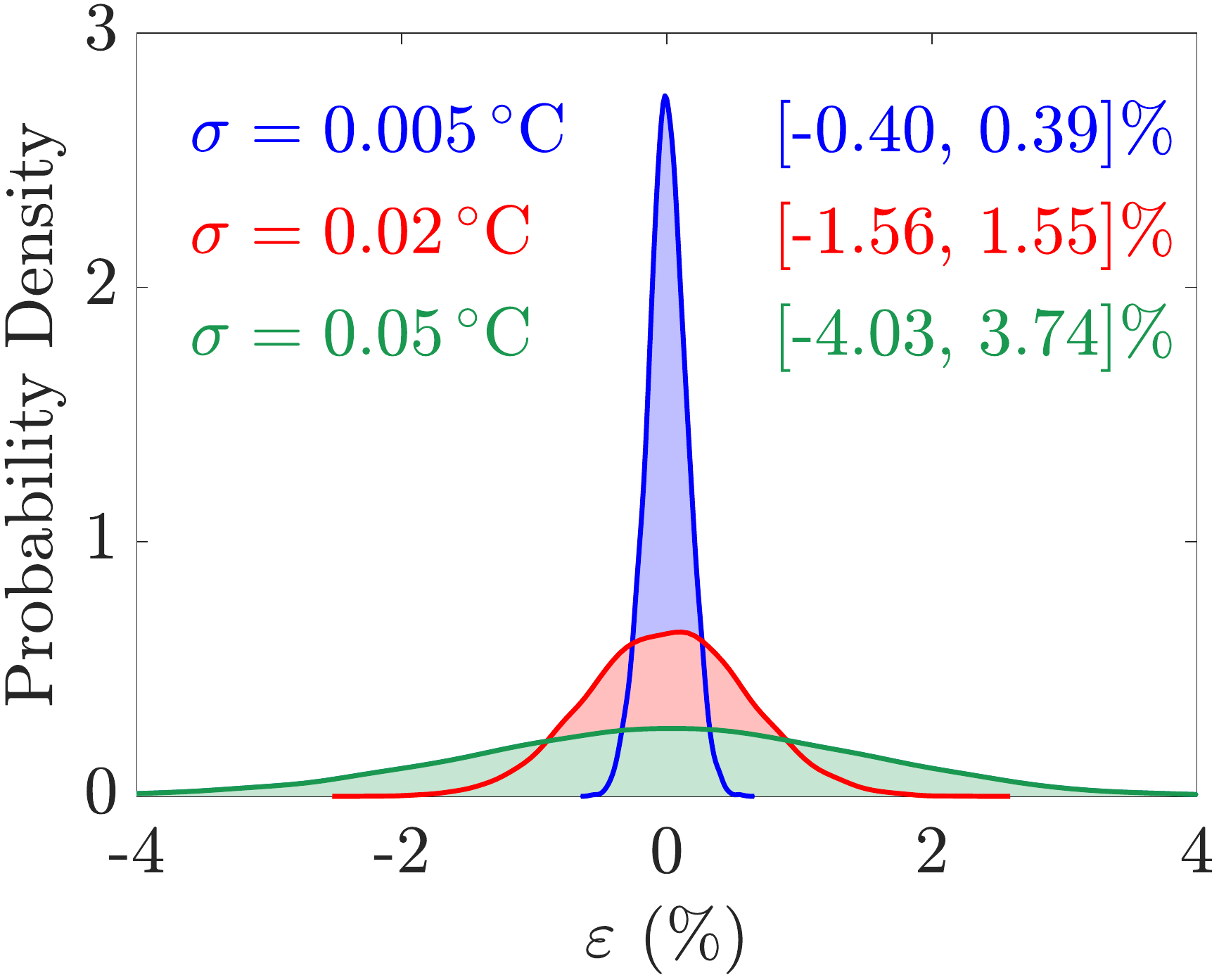}}\hfill\subfloat[]{\includegraphics[width=0.31\textwidth]{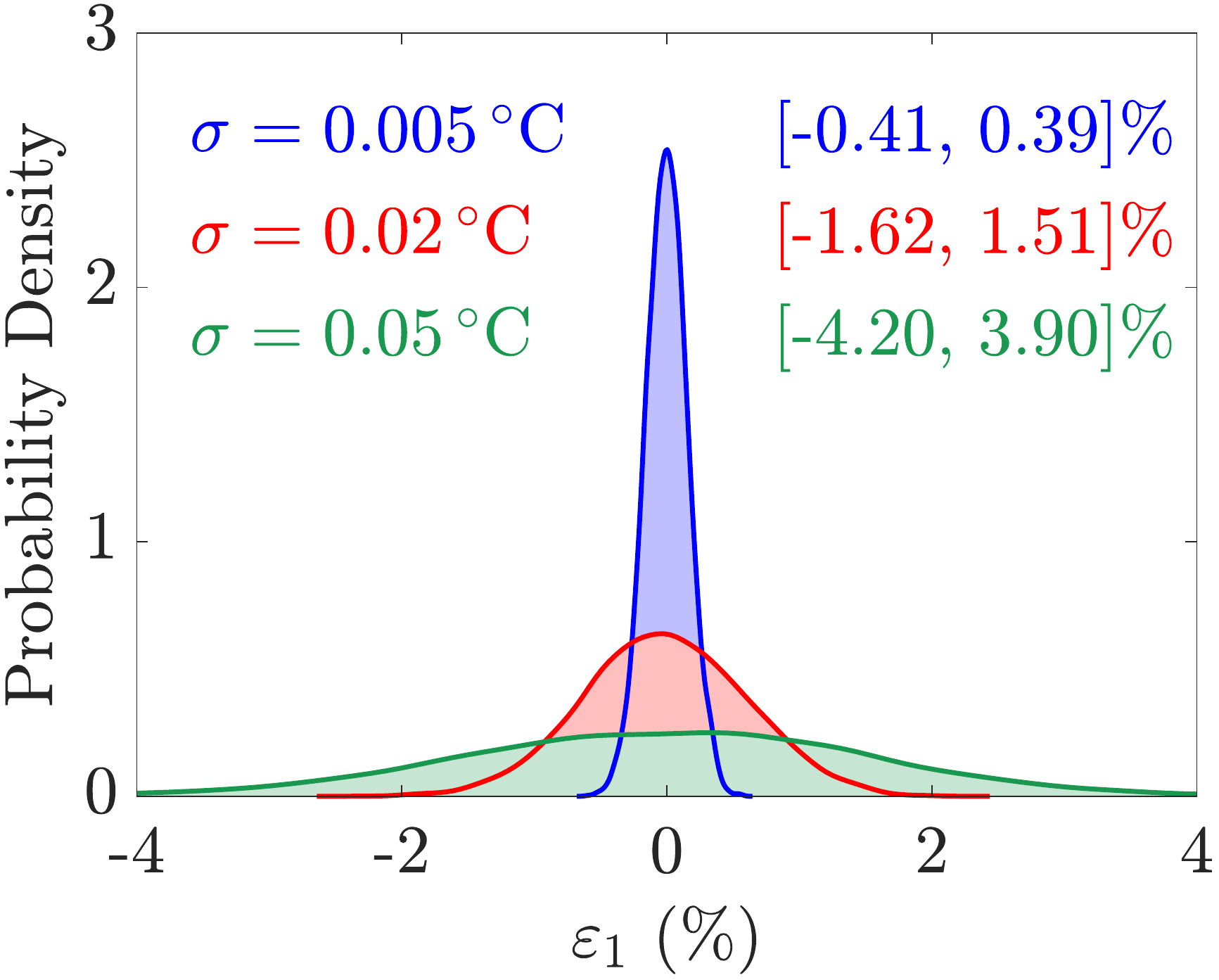}}\hfill\subfloat[]{\includegraphics[width=0.31\textwidth]{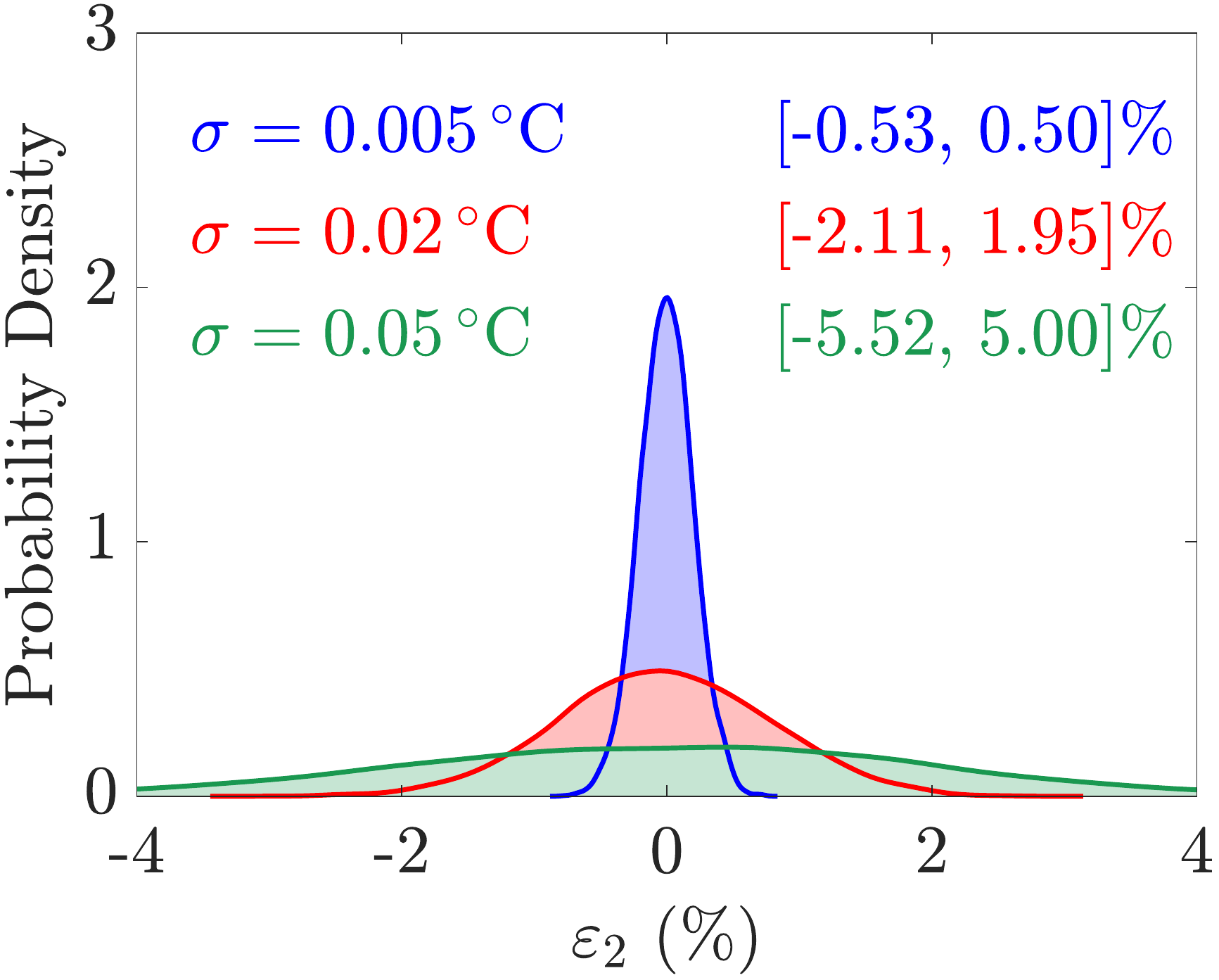}}
\caption{(a)--(c) Example realisations of the rear-surface temperature rise history using low, moderate and high levels of noise. (d)--(f) Signed relative error distributions for the thermal diffusivity estimates (d) $\alpha$ (\ref{eq:model1_alpha}) (e) $\alpha_{1}$ (\ref{eq:model2_alpha1}) and (f) $\alpha_{2}$ (\ref{eq:model2_alpha2}) each implemented using the numerical approximation of $I_{T}$ (\ref{eq:IT_approx}). Distributions and [0.5\%,99.5\%] quantile intervals are given for each of the three noise levels, with each distribution constructed using 10,000 realisations. All results are for the parameter values in Table \ref{tab:parameter_values} and the exponential pulse (\ref{eq:exponential_pulse}) with $\beta = 0.001\,\mathrm{s}$.}
\label{fig:results1}
\end{figure*}

In Figure \ref{fig:results1}, results are reported for 10,000 realisations, low ($\sigma = 0.005\,^{\circ}\mathrm{C}$), moderate ($\sigma = 0.02\,^{\circ}\mathrm{C}$) and high ($\sigma = 0.05\,^{\circ}\mathrm{C}$) levels of noise, and the exponential pulse (\ref{eq:exponential_pulse}) with $\beta = 0.001\,\mathrm{s}$ (as considered in the previous section). Example rear-surface temperature rise histories for each level of noise are given in Figures \ref{fig:results1}(a)--(c). The relative error distributions in Figures \ref{fig:results1}(d)--(f) are smoothed using MATLAB's inbuilt kernel smoothing function \texttt{ksdensity}. These figures also include the $0.5\%$ and $99.5\%$ quantiles for the data; for example, 99\% of the estimates of $\alpha_{1}$ have a relative error of between $-1.62\%$ and $+1.51\%$ for $\sigma = 0.02\,^{\circ}\mathrm{C}$ (Figure \ref{fig:results1}e). Similar results are obtained for the rectangular and triangular pulses and are therefore not reported. Comparing Figures \ref{fig:results1}(e)--(f), we see that the thermal diffusivity estimates are slightly less accurate for the second layer than the first layer. This is because for this particular problem $\alpha_{1}>\alpha_{2}$ (Table \ref{tab:parameter_values}) so the small error incurred through approximating $I_{T}$ via the trapezoidal rule (\ref{eq:IT_approx}) is amplified by a greater amount in Eq (\ref{eq:model2_alpha2}), where it is multiplied by $\alpha_{1}$, than in Eq (\ref{eq:model2_alpha1}), where it is multiplied by $\alpha_{2}$. Overall, the results in Figures \ref{fig:results1}(d)--(f) are comparable in accuracy to those reported previously for the rear-surface integral method \cite{carr_2019}.

\begin{figure*}[ht]
\centering
\subfloat[]{\includegraphics[width=0.32\textwidth]{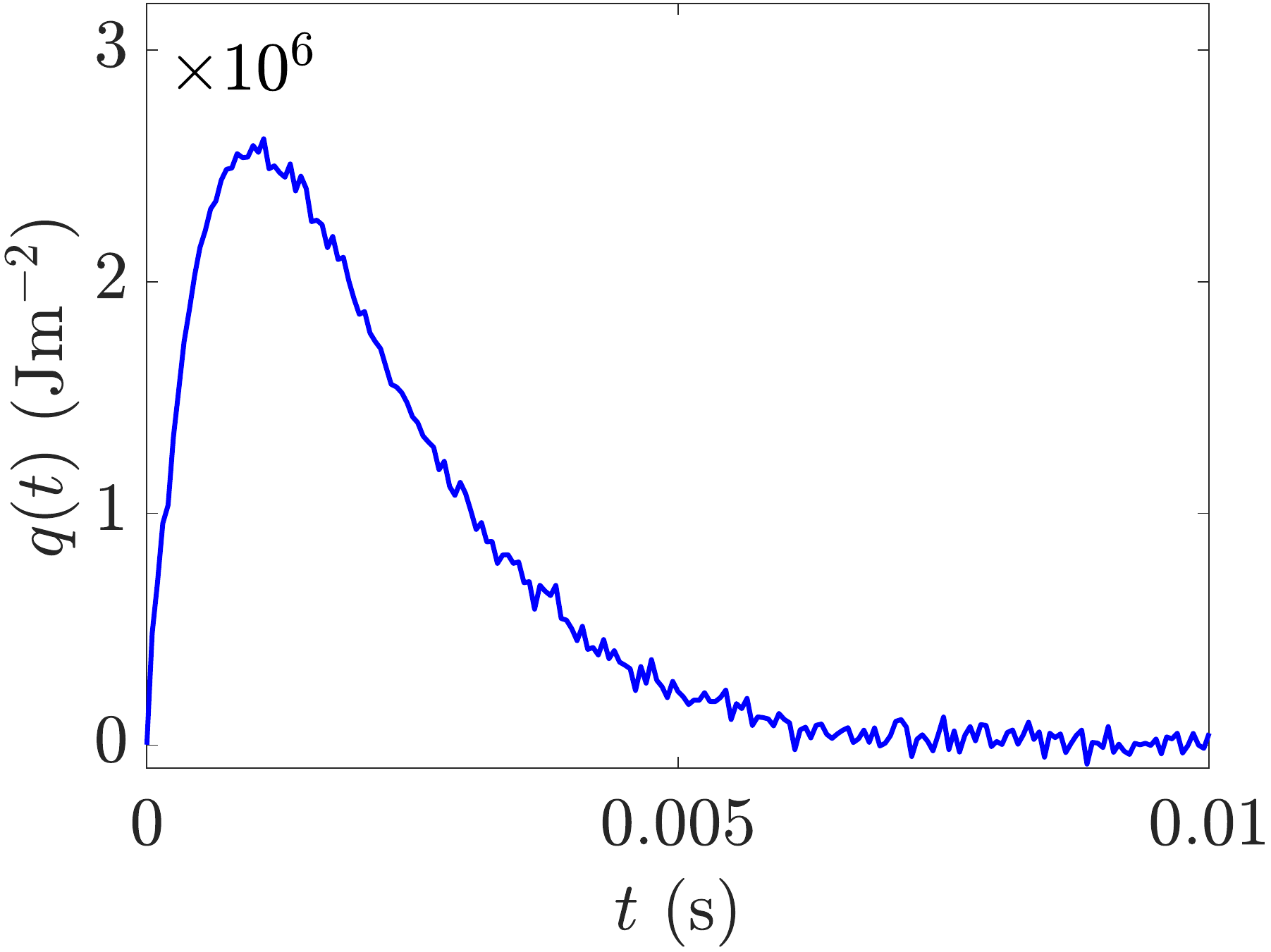}}\hspace{0.04\textwidth}\subfloat[]{\includegraphics[width=0.31\textwidth]{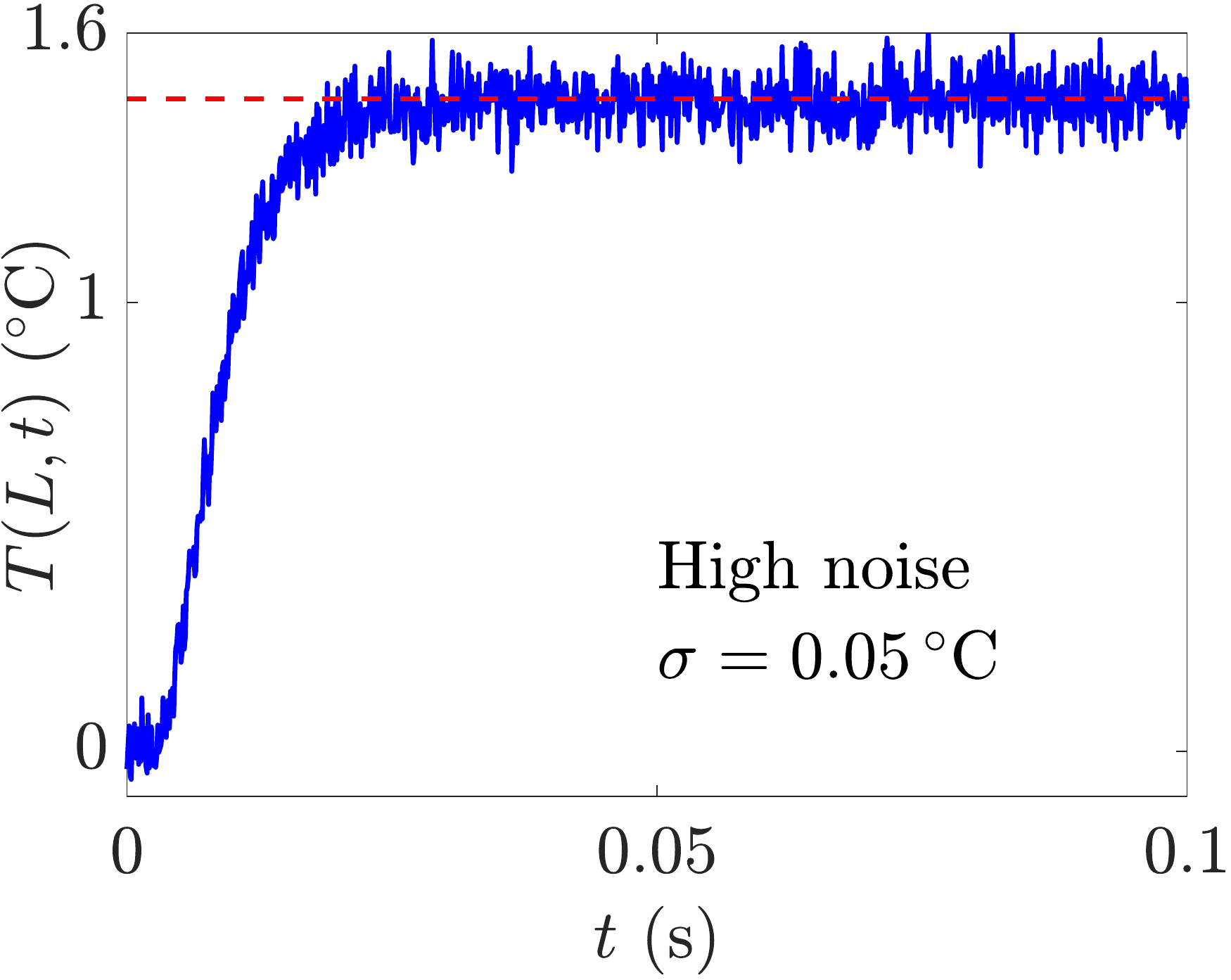}}\\
\subfloat[]{\includegraphics[width=0.31\textwidth]{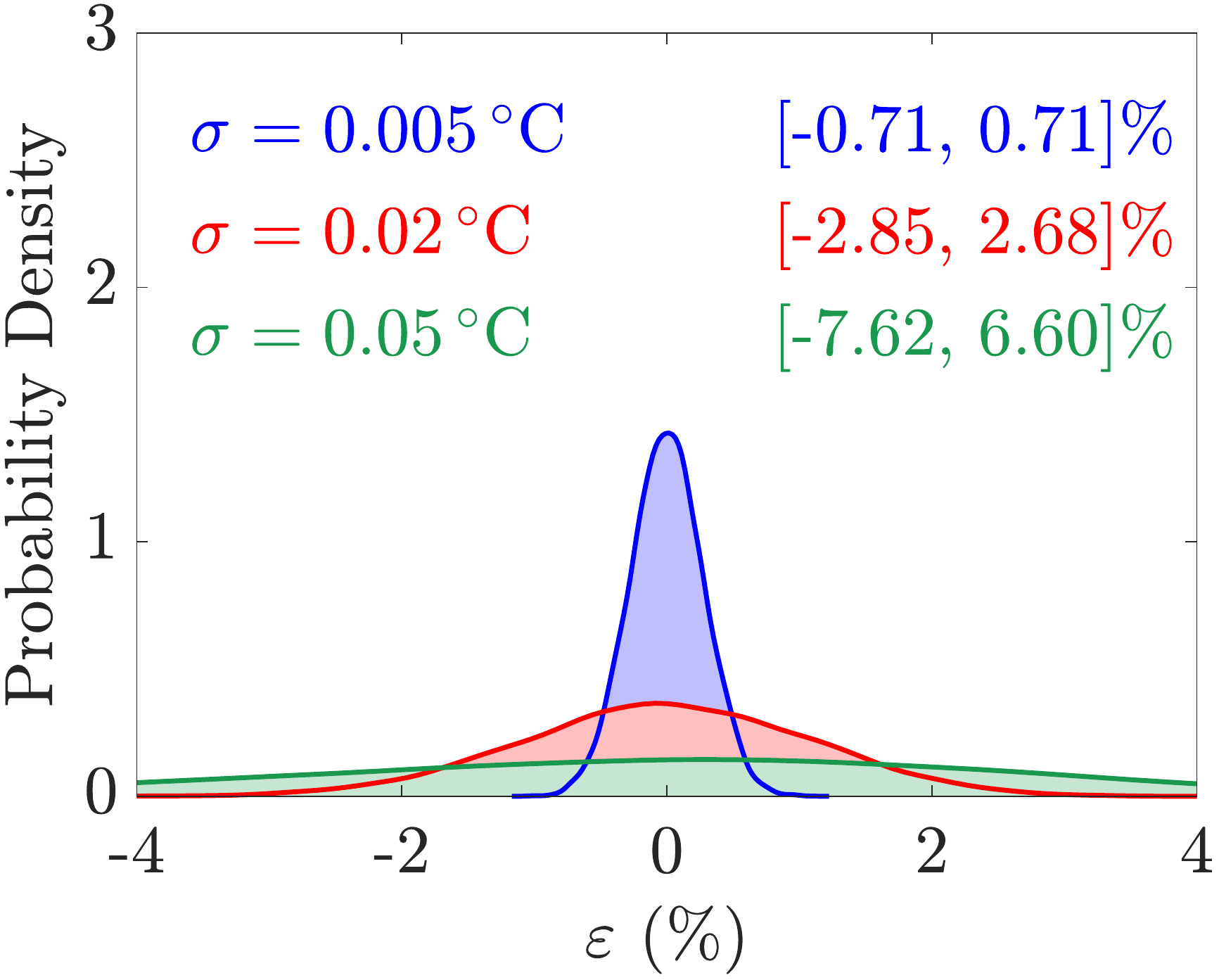}}\hfill
\subfloat[]{\includegraphics[width=0.31\textwidth]{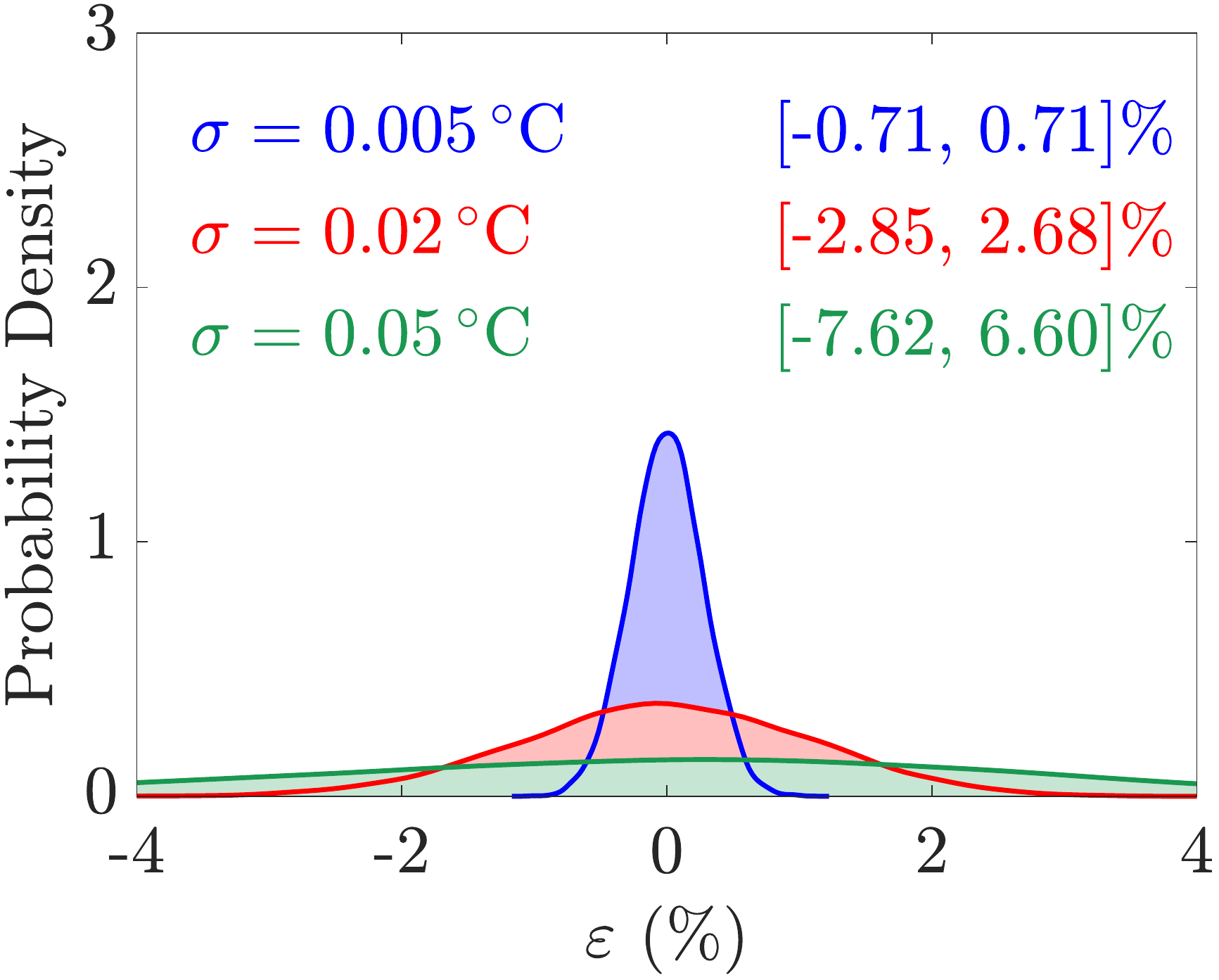}}\hfill\subfloat[]{\includegraphics[width=0.31\textwidth]{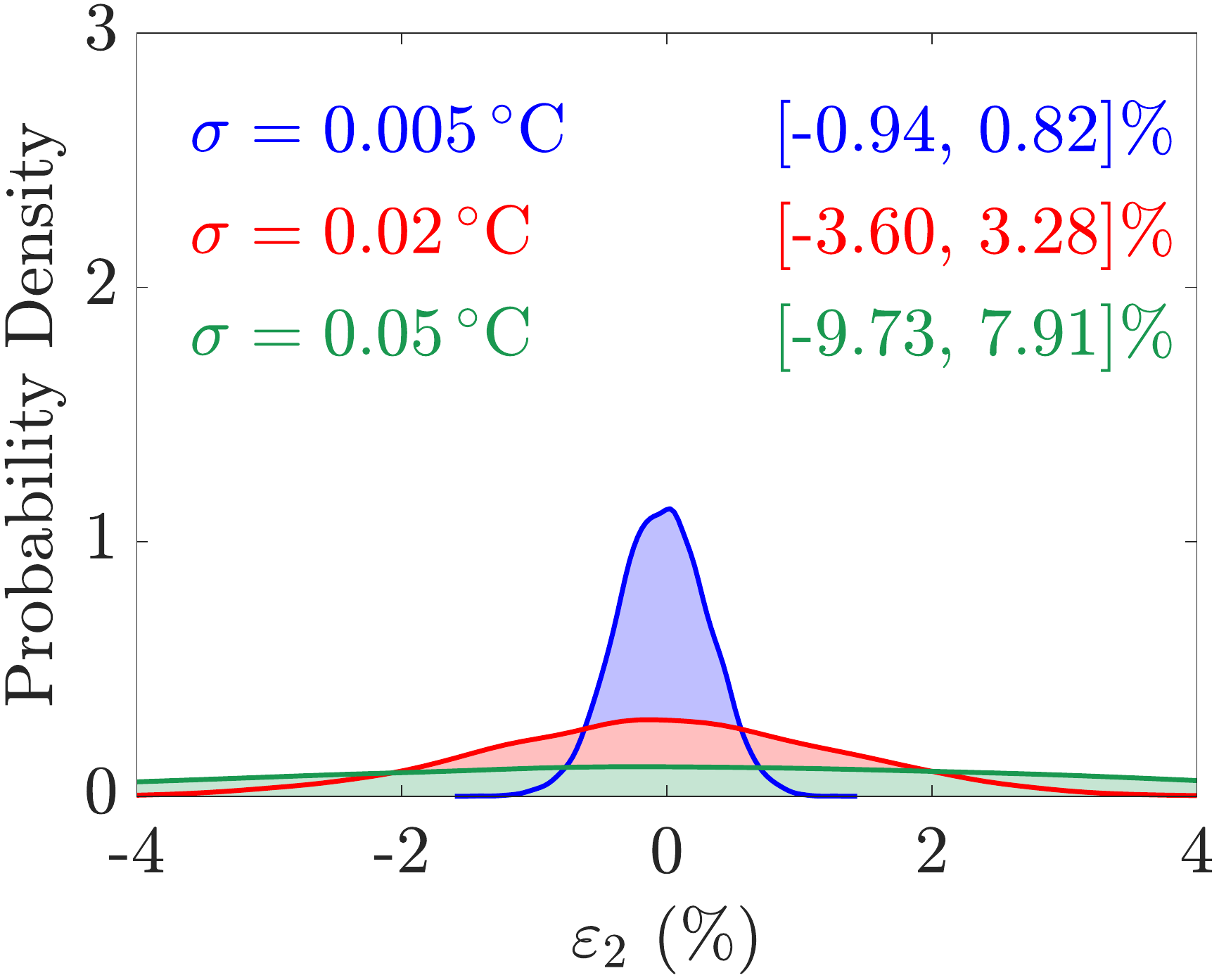}}
\caption{(c)--(e) Signed relative error distributions for the thermal diffusivity estimates (c) $\alpha$ (\ref{eq:model1_alpha}) (d) $\alpha_{1}$ (\ref{eq:model2_alpha1}) and (e) $\alpha_{2}$ (\ref{eq:model2_alpha2}) each implemented using the numerical approximations of $I_{T}$ (\ref{eq:IT_approx}), $Q_{\infty}$ (\ref{eq:Q_approx}), $I_{q}$ (\ref{eq:Iq_approx}) and $T_{\infty}$ (\ref{eq:Tinf_approx}). Distributions and [0.5\%,99.5\%] quantile intervals are given for each of the three noise levels, with each distribution constructed using 10,000 realisations; (b) provides an example realization. All results are for the parameter values in Table \ref{tab:parameter_values} and the noisy exponential pulse (\ref{eq:exponential_pulse}) with $\beta = 0.001\,\mathrm{s}$ shown (a).}
\label{fig:results2}
\end{figure*}

Our numerical experiments so far have assumed that the laser pulse shape can be described exactly by either the rectangular (\ref{eq:rectangular_pulse}), triangular (\ref{eq:triangular_pulse}) or exponential (\ref{eq:exponential_pulse}) pulse. We have also calculated the steady-state temperature, $T_{\infty}$, assuming known values for the amount of heat absorbed at the front surface, $Q_{\infty}$, and the volumetric heat capacity(s), $\rho c$, $\rho_{1}c_{1}$ and $\rho_{2}c_{2}$ (i.e., by evaluating Eq (\ref{eq:model1_Tinf}) and Eq (\ref{eq:model2_Tinf}) directly using the parameter values in Table \ref{tab:parameter_values}). We now explore calculating the thermal diffusivity directly using the discrete pulse shape measurements that arise in a laser flash experiment and a value of $T_{\infty}$ inferred from the rear-surface temperature rise history ($\widetilde{T}_{i}$ where $i = 0,\hdots,N$). Given discrete values of the heat pulse over time, $q(\hat{t}_{j})$ where $\hat{t}_{j} = j\delta t/M$ and $j = 0,\hdots,M$ (note that the time between pulse measurements $\delta t$ differs from the time between rear-surface temperature rise measurements $\Delta t$), we approximate $Q_{\infty}$ using the trapezoidal rule:
\begin{gather}
\label{eq:Q_approx}
Q_{\infty} \approx \frac{\delta t}{2}\sum_{j=1}^{M} \bigl[q(\hat{t}_{j-1})+q(\hat{t}_{j})\bigr],
\end{gather}
and $I_{q}$ using two applications of the trapezoidal rule
\begin{align}
I_{q} &\approx \Delta t \sum_{i=1}^{N}\left[1 - \frac{1}{Q_{\infty}}\left(\int_{0}^{\hat{t}_{i-1}}q(s)\,\text{d}s + \int_{0}^{\hat{t}_{i}}q(s)\,\text{d}s\right)\right],\\
\nonumber
&\approx \Delta t\sum_{i=1}^{N}\left[1 - \frac{\delta t}{4Q_{\infty}}\left(\sum_{j=1}^{N_{i-1}}\bigl[q(\hat{t}_{j-1}) + q(\hat{t}_{j})\bigr]\right.\right.\\\label{eq:Iq_approx}
 &\hspace{0.198\textwidth} \left.\left.+ \sum_{j=1}^{N_{i}} \bigl[q(\hat{t}_{j-1}) + q(\hat{t}_{j})\bigr]\right)\right],
\end{align}
where $N_{i} = t_{i}/\delta t$ is the number of intervals of length $\delta t$ comprising the interval $[0,t_{i}]$ (i.e., the approximation of $I_{q}$ assumes that $\Delta t/\delta t$ is an integer). To approximate $T_{\infty}$ from the rear-surface temperature rise history, we use the height of the horizontal line of best fit through the last $p$ measurements (Figure \ref{fig:results2}b), which yields the mean value 
\begin{gather}
\label{eq:Tinf_approx}
T_{\infty} \approx \frac{1}{p}\sum_{i=1}^{p} \widetilde{T}_{N-p+i}.
\end{gather}
Results in Figures \ref{fig:results2}(c)--(e) are given for the discrete heat pulse depicted in Figure \ref{fig:results2}(a), consisting of 201 discrete values between $t = 0$ and $t = 0.01\,\text{s}$ (inclusive) giving $M = 200$ and $\delta t = 5\times 10^{-5}\,\text{s}$ in the approximations of $Q_{\infty}$ (\ref{eq:Q_approx}) and $I_{q}$ (\ref{eq:Iq_approx}). The discrete values are randomly generated by adding Gaussian noise of mean zero and standard deviation $40 000\,\text{J}\text{m}^{-2}$ to the exponential pulse (\ref{eq:exponential_pulse}) with $\beta = 0.001\,\mathrm{s}$. Figure \ref{fig:results2}(b) provides an example realisation of the rear-surface temperature rise history for the homogeneous sample obtained by adding Gaussian noise to the solution of the heat flow model (\ref{eq:model1_pde})--(\ref{eq:model1_bcs}) with $q(t)$ defined as in Figure \ref{fig:results2}(a). The horizontal red dashed line in Figure \ref{fig:results2}(b) depicts the utilised approximation of $T_{\infty}$ (\ref{eq:Tinf_approx}) with $p = 200$. Comparing Figures \ref{fig:results1}(d)--(f) and Figures \ref{fig:results2}(c)--(e) we see that the approximations (\ref{eq:Q_approx}) and (\ref{eq:Iq_approx}) roughly double the length of the [0.5\%, 99.5\%] quantile interval.  

\section{Conclusion}
\label{sec:conclusions}
This paper has focussed on methods for calculating thermal diffusivity from laser flash experimental data. Specifically, we have shown how to account for real heat pulse shapes and two-layer samples using the recently-proposed rear-surface integral method \cite{carr_2019}. The main contributions of the paper are new formulas for calculating the thermal diffusivity that correct for arbitrary heat pulse shapes:
\begin{enumerate}[(i)]
\item Eqs (\ref{eq:model1_alpha})--(\ref{eq:model1_Iq}) for calculating the thermal diffusivity of a homogeneous sample; and
\item Eqs (\ref{eq:model2_alpha1})--(\ref{eq:model2_Iq}) for calculating the thermal diffusivity of one layer of a two-layer sample given the thermal diffusivity of the other layer. 
\end{enumerate}
Computational results confirmed the accuracy of the derived formulas and demonstrated how the formulas can be applied to the kinds of experimental data arising from the laser flash experiment.

The formulas derived in this paper are limited to thermally insulated samples, in which case the heat flow model reduces to a one-dimensional model in the direction of the thickness of the sample. \revision{Incorporating heat losses from the front and rear surfaces of the sample can be carried out by combining the analysis in this paper with that presented previously by \citet[Section 5]{carr_2019}.} Possible directions for future research also include extension of the rear-surface integral method with finite-pulse time effects to three-layer samples or two or three dimensional heat flow models; the latter of which could allow for heat losses from the other surfaces of the sample (Figure \ref{fig:laser_flash}b--c) to be accounted for in the formulas.

The analysis presented in this paper can also be extended to other applications involving diffusion in homogeneous or layered materials in a straightforward manner. For example, by appropriately modifying the boundary and interface conditions (\ref{eq:model2_bc1})--(\ref{eq:model2_intc2}), the rear-surface integral method can help parameterise a mathematical model of heat transfer through layered skin formulated by \citet{mcinerney_2019}.

\section*{Acknowledgements}
This research was partially supported by the Australian Mathematical Sciences Institute (AMSI) who provided CJW with a 2018-2019 Vacation Research Scholarship.

\appendix
\section{Homogeneous finite volume scheme}
\label{app:single_layer}
\revision{In this appendix, we outline the finite volume scheme used to obtain a numerical solution to the homogeneous heat flow model (\ref{eq:model1_pde})--(\ref{eq:model1_bcs}). We discretise the interval $[0,L]$ using a uniform grid consisting of $n$ nodes such that the $k$th node is located at $x = kh =: x_{k}$, where $k = 1,\hdots,n$ and $h = L/(n-1)$. Let $T_{k}(t)$ denote the numerical approximation to $T(x,t)$ at $x = x_{k}$. The finite volume scheme takes the form:
\begin{align*}
\frac{\text{d}T_{k}}{\text{d}t} = G_{k},\quad k = 1,\hdots,n,
\end{align*}
with initial condition $T_{k} = 0$ at $t = 0$ (\ref{eq:model1_ic}) for $k = 1,\hdots,n$ and right-hand side defined by
\begin{gather*}
G_{k} = 
\begin{cases}
\displaystyle
\frac{2\alpha(T_{k+1}-T_{k})}{h^{2}} + \frac{2q(t)}{h\rho c},\;\; k = 1,\\[0.4cm]
\displaystyle
\frac{\alpha(T_{k+1}-2T_{k}+T_{k-1})}{h^{2}},\;\; k = 2,\hdots,n-1,\\[0.4cm]
\displaystyle
\frac{2\alpha(T_{k-1}-T_{k})}{h^{2}},\;\; k = n.
\end{cases}
\end{gather*}
This initial value problem is solved using MATLAB 2017B's \texttt{ode15s} solver on the interval $[0,t_{N}]$ with specified absolute and relative tolerances (\texttt{AbsTol} and \texttt{RelTol}) of $10^{-12}$ and the interval of integration ($\texttt{tspan}$) specified to return the solution at $N+1$ equally-spaced discrete times: $t_{i} = i\Delta t$, where $i = 0,\hdots,N$ and $\Delta t = t_{N}/N$. Using this solution, the rear-surface temperature rise history is given by the discrete values $T_{n}(t_{0}),\hdots,T_{n}(t_{N})$.}

\section{Two-layer finite volume scheme}
\label{app:two_layer}
\revision{In this appendix, we describe the finite volume scheme used to solve the two-layer heat flow model (\ref{eq:model2_pde1})--(\ref{eq:model2_intc2}). We discretise the interval $[0,L]$ using a grid consisting of $n = n_{1}+n_{2}-1$ nodes such that the $k$th node is located at $x = x_{k}$ with $x_{k} = (k-1)h_{1}$ for $k = 1,\hdots,n_{1}$ and $x_{k} = \ell_{1}+(k-n_{1})h_{2}$ for $k = n_{1}+1,\hdots,n_{1}+n_{2}-1$, where $h_{1} = \ell_{1}/(n_{1}-1)$ and $h_{2} = \ell_{2}/(n_{2}-1)$. Let $T_{k}(t)$ approximate $T(x_{k},t)$ as in  \ref{app:single_layer}. The finite volume scheme takes the form:
\begin{align*}
\frac{\text{d}T_{k}}{\text{d}t} = G_{k},\quad k = 1,\hdots,n_{1}+n_{2}-1,
\end{align*}
with initial condition $T_{k} = 0$ at $t = 0$ (\ref{eq:model2_ic1})--(\ref{eq:model2_ic2}) for $k = 1,\hdots,n_{1}+n_{2}-1$ and right-hand side defined by
\begin{gather*}
\hspace*{-0.1cm}G_{k} = 
\begin{cases}
\displaystyle
\frac{2\alpha_{1}(T_{2}-T_{1})}{h_{1}^{2}} + \frac{2q(t)}{h_{1}\rho_{1} c_{1}},\;\; k = 1,\\[0.4cm]
\displaystyle
\frac{\alpha_{1}(T_{k+1}-2T_{k}+T_{k-1})}{h_{1}^{2}},\;\; k = 2,\hdots,n_{1}-1,\\[0.4cm]
\displaystyle
\frac{2\left[\frac{k_{2}}{h_{2}}T_{k+1} - \left(\frac{k_{1}}{h_{1}} + \frac{k_{2}}{h_{2}}\right)T_{k} + \frac{k_{1}}{h_{1}} T_{k-1}\right]}{\rho_{1}c_{1}h_{1} + \rho_{2}c_{2}h_{2}},\;\; k = n_{1},\\[0.5cm]
\displaystyle
\frac{\alpha_{2}(T_{k+1}-2T_{k}+T_{k-1})}{h_{2}^{2}},\;\; k = n_{1}+1,\hdots,n_{1}+n_{2}-2,\\[0.4cm]
\displaystyle
\frac{2\alpha_{2}(T_{k-1}-T_{k})}{h_{2}^{2}},\;\; k = n_{1}+n_{2}-1.
\end{cases}
\end{gather*}
This initial value problem is then solved to obtain the rear-surface temperature rise history in an identical manner to that described for the homogeneous case at the end of \ref{app:single_layer}.}

%\section*{References}

%\bibliographystyle{model2-names}
%\bibliographystyle{elsarticle-num}
\bibliographystyle{model1-num-names}
\bibliography{references}

\end{document}